\documentclass{ws-ijmpd}

\begin{document}

\markboth{R. Maier and I. D. Soares}
{Gravitational Collapse and Black Hole Thermodynamics in Braneworld Scenario}

\catchline{}{}{}{}{}
%
%%%%%%%%%%%%%%%%%%%%%%%%%%%%%%%%%%%%%%%%%%%%%%%%%%%%%%%%%%%%%%%%%%%%

\title{\bf{GRAVITATIONAL COLLAPSE AND BLACK HOLE THERMODYNAMICS IN BRANEWORLD SCENARIO}}

\author{R. MAIER\footnote{E-mail address: rodmaier@cbpf.br} ~and I. DAMI\~AO SOARES\footnote{E-mail address: ivano@cbpf.br}}

\address{Centro Brasileiro de Pesquisas F\'isicas,\\
Rua Dr. Xavier Sigaud 150, Urca,\\
Rio de Janeiro, CEP 22290-180-RJ, Brazil}

\maketitle

\begin{history}
\received{Day Month Year}
\revised{Day Month Year}
\comby{Managing Editor}
\end{history}

\begin{abstract}
We examine the dynamics of the gravitational collapse in a 4-dim Lorentzian brane
embedded in a 5-dim bulk with an extra timelike dimension. By considering the collapse
of pure dust on the brane we derive a bouncing FLRW interior solution and match it
with a corrected Schwarzschild exterior geometry. In the physical domain considered for
the parameters of the solution, the analytical extension is built, exhibiting an exterior
event horizon and a Cauchy horizon, analogous to the Reissner-Nordstr\"om solution. For
such an exterior geometry we examine the effects of the bulk-brane corrections in the
Hawking radiation. In this scenario the model extends Bekenstein's black hole geometrical
thermodynamics for quasi-extremal configurations, with an extra work term in the
laws associated with variations of the brane tension. We also propose a simple statistical
mechanics model for the entropy of the bouncing collapsed matter by quantizing
its fluctuations and constructing the associated partition function. This entropy differs
from the geometrical entropy by an additive constant proportional to the area of the
extremal black hole and satisfies an analogous first law of thermodynamics. A possible
connection between both entropies is discussed.
\end{abstract}


\section{Introduction}	
Black holes are solutions of vacuum general relativity equations describing the exterior
spacetime of the final stage of gravitationally bounded systems whose masses
exceeded the limits for a finite equilibrium configuration.\cite{chandra} Geometrically a black
hole may be described as a region of asymptotically flat spacetimes bounded by an
event horizon hiding a singularity formed in the collapse. Fundamental theorems
by Israel and Carter\cite{carter2,carter3} state that the final stage of a general collapse of uncharged
matter is typically a Kerr black hole, which has an involved singularity structure.
\par
Nevertheless, for a realistic gravitational collapse we have no evidence that the
Kerr solution describes accurately the interior geometry of the black hole. On the
contrary, the best theoretical evidence presently available indicates that the interior
of the black hole thus formed is analogous to the interior of a Schwarzschild
black hole with a global spacelike singularity.\cite{wald} The simplest way of forming such
structure is by the spherical collapse of dust, as originally shown in the classical
paper of Oppenheimer and Snyder.\cite{oppen} However, as singularities cannot be empirically
conceived, this turns out to be a huge pathology of the theory.
\par
Notwithstanding the cosmic censorship hypothesis\cite{wald1} (CCH), there is no doubt
that the general theory of relativity must be properly corrected or even replaced
by a completely new theory, let us say a quantum theory of gravity. This demand
is in order to solve the issue of the presence of singularities predicted by classical
general relativity, either in the formation of a black hole or in the beginning of the
universe. While a full quantum gravity theory remains presently an elusive theoretical
problem, quantum gravity corrections near singularities formed by gravitational
collapse have been the object of much recent research, from loop quantum cosmology\cite{bojowald}
to D-brane theory.\cite{rs}\cdash\cite{maartens2} In the latter scenario extra dimensions are introduced
constituting the bulk space. All matter would be trapped on a 4-dim world-brane
spacetime embedded in the bulk and only gravitons would be allowed to move in
the full bulk. At low energies general relativity is recovered\cite{rs} but at high energy
scales significant changes are introduced into the gravitational dynamics and the
singularity could be eventually removed.
\par 
The problem of the gravitational collapse in the braneworld scenario has been
the object of several important works. Bruni {\it et al.}\cite{bruni} studied the Oppenheimer–
Snyder collapse on a Randall-Sundrum-type brane and showed that the exterior
vacuum spacetime on the brane cannot be static and therefore precluding the formation
of black holes in the theory. However Dadhich {\it et al.}\cite{dadhich} have demonstrated
the existence of static black holes on the brane in the Randall-Sundrum scenario.
These black holes are exact solutions of the effective Einstein equations on the
brane and correspond to Reissner–N\"ordstrom (RN)-type black holes, with a {\it tidal
charge} (also denoted Kaluza–Klein (KK) charge) originated from the 5-dim Weyl
curvature instead of an electric charge. In this vein, Govender and Dadhich\cite{dadhich1} constructed
a model of the Oppenheimer–Snyder collapse in the brane in which the
collapsing solution is matched to the brane generalized Vaydia solution which in
turn is matched to the asymptotically flat RN-type metric with a KK {\it charge}.
The mediation by the Vaydia radiation metric is a new feature introduced by the
Randall-Sundrum-type brane so that the collapsing sphere radiates null radiation.
This picture is the paradigm of the gravitational collapse of a homogeneous spherically
symmetric configuration on a Randall-Sundrum-type brane embedded in a
nonconformally flat, but otherwise vacuum bulk. The problem of the gravitational
collapse of a null fluid on the brane was approached by Dadhich and Ghosh,\cite{dadhich2} where
the parameter windows in the initial data set, giving rise to a naked singularity or
favoring the formation of black holes, are examined.
\par
In our approach in this paper we have considered the gravitational collapse of a spherically symmetric dust distribution
in the framework of a braneworld scenario, with a 5D bulk having an extra timelike dimension. The interior geometry is still
given by the FLRW metric but -- due to the timelike character of the extra dimension -- the dynamics of the collapsing
dust has an effective potential barrier generated by the bulk-brane corrections. This potential barrier avoids the
formation of a singularity yielding a perpetually oscillating collapsed matter. We obtain the unique static exterior
geometry which is smoothly matched to the interior FLRW geometry.
\par
As we know from General Relativity, the CCH addresses the issue whether a singularity thus formed in gravitational collapse is visible to an asymptotic observer or hidden by an event horizon.
As we shall see in our model, if the total mass of the collapsing dust is larger
than a critical value, we obtain a static solution which corresponds to a nonsingular black hole with an event horizon (besides a Cauchy horizon) encapsulating not a singularity but a perpetually oscillating collapsed matter.
Although this perpetually bouncing matter is not visible from an asymptotic observer, no singularity is engendered.
This new feature gives rise to a modified CCH which now addresses the issue whether the perpetually bouncing matter is visible to asymptotic observer or hidden by an event horizon.
\par
By considering our exterior static solution, we construct a statistical model
for the quantum degrees of freedom of the oscillating collapsed matter whose entropy can be associated with the
entropy fluctuations about the extremal configuration of the exterior static black hole geometry.
In this direction we are also led to evaluate the Hawking evaporation processes of the exterior black hole.
The no-go theorem of Ref. 14 may be circumvented as long as one relaxes the condition of a vacuum nonconformally flat bulk assuming, for instance, a bulk matter content satisfying energy conditions
or the presence of torsion degrees of freedom in the bulk.\cite{maier} We have however not addressed the 5D
equations for the determination of the bulk space since this task is beyond the scope of this paper.
\par
For the sake of completeness let us give a brief introduction to braneworld theory,
making explicit the specific assumptions used in
obtaining the dynamics of the model. We rely on Refs.
9-12, and our notation
basically follows \cite{wald}. Let us start with a 4-dim Lorentzian
brane $\Sigma$ with metric $g_{ab}$, embedded in a 5-dim conformally flat bulk $\cal{M}$ with
metric $g_{AB}$. Capital Latin indices range from 0 to 4, small Latin
indices range from 0 to 3. We regard $\Sigma$ as a common boundary of two pieces
${\cal{M}}_{1}$ and ${\cal{M}}_{2}$ of $\cal{M}$ and the metric $g_{ab}$ induced on the brane by the metric of the two pieces should coincide although the extrinsic
curvatures of $\Sigma$ in ${\cal{M}}_{1}$ and ${\cal{M}}_{2}$ are allowed to be different.
The action for the theory has the general form
\begin{eqnarray}
\label{eq4r}
\nonumber
S=\frac{1}{2\kappa^2_{5}}\Big\{\int_{M_{1}}\sqrt{-\epsilon ~^{(5)}g} \Big[^{(5)}R-2\Lambda_{5} +2\kappa^2_{5}L_{5}\Big]d^5x~~~~~~~~~~~~~~~~~~~~~~~~~~~~~~~~~~\\
\nonumber
+\int_{M_{2}}\sqrt{-\epsilon ~^{(5)}g} \Big[^{(5)}R-2\Lambda_{5}+2\kappa^2_{5}L_{5}\Big]d^5x
\nonumber
+2\epsilon\int_{\Sigma}\sqrt{- ^{(4)}g}K_{2}d^4x\\-2\epsilon\int_{\Sigma}\sqrt{-^{(4)}g}K_{1}d^4x\Big\}
\nonumber
+\frac{1}{2}\int_{\Sigma}\sqrt{-^{(4)}g}\Big(\frac{1}{2\kappa^2_{4}}^{(4)}R-2\sigma\Big)d^4x\\
~~~~~~~~~~~~~~~~+\int_{\Sigma}\sqrt{-^{(4)}g} L_{4}(g_{\alpha\beta},\rho)d^4x.
\end{eqnarray}
In the above $^{(5)}R$ is the Ricci scalar of the Lorentzian 5-dim metric $g_{AB}$
on $\cal{M}$, and $^{(4)}R$ is the scalar curvature of the induced metric $g_{ab}$
on $\Sigma$. The parameter $\sigma$ is denoted the brane tension. The unit vector $n^{A}$ normal to the boundary $\Sigma$ has norm $\epsilon$. If $\epsilon=-1$ the
signature of the bulk space is $(-,-,+,+,+)$, so that the extra dimension is timelike.
The quantity $K=K_{ab}~ g^{ab}$ is the trace of the
symmetric tensor of extrinsic curvature
$K_{ab}= Y_{,a}~^{C}~Y_{,b}~^{D}~{\nabla_{C}}{n_{D}}$, where $Y^{A}(x^a)$ are the
embedding functions of $\Sigma$ in $\cal{M}$\cite{eisenhart}. While
$L_{4} (g_{ab}, \rho)$
is the Lagrangean density of the perfect fluid\cite{taub}(with equation of state $p= \alpha\rho$),
whose dynamics is restricted to the brane $\Sigma$, $L_{5}$ denotes the lagrangian of matter in the bulk. All integrations over the bulk and the
brane are taken with the natural volume elements $\sqrt{-\epsilon~ {^{(5)}}g}~ d^{5}x$ and $\sqrt{- {^{(4)}}g}~ d^{4}x$ respectively. $\kappa_{5}$ and $\kappa_{4}$ are Einstein constants in five and four-dimensions. With the exception of Section 6, throughout the paper we use units such that $\hbar=c=1$.
\par Variations that leave the induced metric on $\Sigma$
intact result in the equations
\begin{eqnarray}
\label{eq6r}
^{(5)}G_{AB}+ \Lambda_5~{^{(5)}}g_{AB}=\kappa^2_5 {^{(5)}}T_{AB},
\end{eqnarray}
while considering arbitrary variations of $g_{AB}$ and taking into account (\ref{eq6r})
we obtain
\begin{eqnarray}
\label{eq7r}
^{(4)}G_{ab}+\epsilon ~\frac{\kappa_4}{\kappa_5}\Big( S^{(1)}_{ab}+S^{(2)}_{ab}\Big )
=\kappa_4\Big(\tau_{ab}-\sigma g_{ab} \Big),
\end{eqnarray}
where $S_{ab} \equiv K_{ab}-K g_{ab}$. In the limit
$\kappa_4 \rightarrow \infty$ equation (\ref{eq7r}) reduces to
the Israel-Darmois junction condition\cite{israel}
\begin{eqnarray}
\label{eq9r}
\Big( S^{(1)}_{ab}+S^{(2)}_{ab}\Big )
=\epsilon~ \kappa_5\Big(\tau_{ab}-\sigma g_{ab} \Big)
\end{eqnarray}
We impose the $Z_2$-symmetry\cite{maartens} and use the junction conditions (\ref{eq9r})
to determine the extrinsic curvature on the brane,
\begin{eqnarray}
\label{eq10r}
K_{ab}=-\frac{\epsilon}{2} \kappa_5 \Big[(\tau_{ab}-\frac{1}{3}\tau g_{ab})+\frac{\sigma}{3} g_{ab} \Big].
\end{eqnarray}
\par
Now using Gauss equation
\begin{eqnarray}
\label{gauss}
^{(4)}R_{abcd}=~^{(5)}R_{MNRS} Y^{M}_{,a} Y^{N}_{,b} Y^{R}_{,c} Y^{S}_{,d}+\epsilon \Big(K_{ac}K_{bd}-K_{ad}K_{bc} \Big)
\end{eqnarray}
together with equations (\ref{eq6r})
and (\ref{eq10r}) we arrive at the induced field equations on the brane
\begin{eqnarray}
\label{eq1.2.13}
^{(4)}G_{ab}=-\Lambda_{4}{^{(4)}g}_{ab}+8\pi G_{N}\tau_{ab}+\epsilon\kappa^4_{5}\Pi_{ab}-\epsilon{E}_{ab}+\epsilon
F_{ab}
\end{eqnarray}
where we define
\begin{eqnarray}
\label{eq1.2.14}
\Lambda_{4}~&:=&\frac{1}{2}\kappa^2_{5}\Big(\frac{\Lambda_{5}}{\kappa^2_5}+\frac{1}{6}\epsilon\kappa^2_{5}\sigma^2\Big),~~~~~~~~~~~~~~~~~~~~~~~~~~~~~~~~~\\
G_{N}&:=&\epsilon\frac{\kappa^4_{5}\sigma}{48\pi},~~~~~~~~~~~~~~~~~~~~~~~~~~~~~~~~~~~~~~~~~~~~~~~~~~~~~\\
\Pi_{ab}&:=&-\frac{1}{4}\tau_{a}^{c}\tau_{bc} +\frac{1}{12}\tau\tau_{ab}
+\frac{1}{8}{^{(4)}g}_{ab}\tau^{cd}\tau_{cd}
-\frac{1}{24}\tau^2{^{(4)}g}_{ab},\\
F_{ab}&:=&\frac{2}{3}\kappa^2_{5}\Big\{\epsilon ~{^{(5)}T}_{BD} Y^B_{,a} Y^D_{,b}+ \Big[{^{(5)}T}_{BD} n^{B} n^{D}-\frac{1}{4}\epsilon~{^{(5)}T}
\Big]{^{(4)}g}_{ab}\Big\}.
\end{eqnarray}
$G_N$ is just the Newton's constant on the brane.
Here we remark that the effective 4-dim cosmological constant can be set zero
in the present case of an extra timelike dimension, by properly fixing the bulk cosmological constant as
$\Lambda_{5}=\frac{1}{6}\kappa_{5}^{4} ~\sigma^2$.
It's important to notice that for a 4-dim brane embedded in a conformally flat bulk we have the absence of the conformal tensor projection and $F_{ab}$ in Eq. (\ref{eq1.2.14}). Accordingly Codazzi's equations imply that
\begin{eqnarray}
\label{eq1.2.17}
\nabla_{a} K-\nabla_{b} K^{b}_{a}=
-\frac{1}{2}\epsilon \kappa^2_{5}\nabla_{b}\tau^{b}_{a},
\end{eqnarray}
By imposing that $\nabla_{b}\tau^{b}_{a}=0$ the Codazzi conditions read
\begin{eqnarray}
\label{eq1.2.18}
\nabla^{a}{E}_{ab}=\kappa^4_{5}\nabla^{a}\Pi_{ab}+\nabla^{a}F_{ab}.
\end{eqnarray}
where $\nabla_a$ is the covariant derivative with respect to the induced metric $g_{ab}$.
Equations (\ref{eq1.2.13}) and (\ref{eq1.2.18}) are the dynamical equations of the
gravitational field on the brane.

\section{The Interior Solution and the Exterior Geometry}

We assume a spacetime braneworld model embedded in a 5-dim de Sitter bulk with a timelike extra dimension ($\epsilon=-1$), whose
matter content is a spherically symmetric collapsing dust with density $\rho$. In a coordinate system comoving with dust the interior geometry is still shown to be a Friedmann-Robertson-Walker metric\cite{weinberg}
\begin{eqnarray}
\label{eq1}
ds^2=-dt^2+a^2(t)\Big(\frac{1}{1-kr^2}dr^2+r^2d\Omega^2\Big)~~
\end{eqnarray}
with its dynamics given by the first order modified Friedmann equation
\begin{eqnarray}
\label{eq2}
\dot{a}^2=- k +\frac{8\pi G_{N} E_{0}}{3a}-\frac{4\pi G_{N} E^2_{0}}{3|\sigma| a^4}
\end{eqnarray}
where we made $\Lambda_4=0$ by a proper choice of $\Lambda_5$. $E_0$ is a constant of motion associated with the dust density, $E_0=\rho a^3$, and $\sigma$ is the
negative brane tension so that $G_{N}>0$.
Assuming initial conditions for the collapse $\dot{a}(0)=0$
and $a(0)=1$, we get
\begin{eqnarray}
\label{eqIII.23}
k=\frac{8\pi G_{N}}{3}\Big[E_{0}-\frac{E^2_{0}}{2|\sigma|}\Big].
\end{eqnarray}
By defining
\begin{eqnarray}
\label{potencial1}
p_{a}:=\dot{a}~,~~V(a):=\frac{2\pi G_{N} E^2_{0}}{3|\sigma| a^4}-\frac{4\pi G_{N} E_{0}}{3a}+\frac{k}{2},
\end{eqnarray}
one may infer the Hamiltonian constraint
\begin{eqnarray}
\label{potencial2}
H=\frac{p_{a}^2}{2}+V(a)=0,
\end{eqnarray}
which is equivalent to the first integral (\ref{eq2}). An immediate calculation shows that from this constraint
one can easily obtain the motion equation
\begin{eqnarray}
\label{eqIII.19}
\nonumber
\ddot{a}a+2\dot{a}^2+2k=\frac{4\pi G_{N} E_{0}}{a}.
\end{eqnarray}
\par
From now on we are going to assume $k>0$ and that the potential $V(a)$ has two real positive roots. These assumptions
restrict the domain of the parameters as
\begin{eqnarray}
\label{restricao}
|\sigma|>2E_{0}.
\end{eqnarray}
Therefore we see that $V(a)$ has one extremal located at $\bar{a}=\Big(\frac{2E_{0}}{|\sigma|}\Big)^{\frac{1}{3}}$, and two positive real roots $a_{1}=a_{min}$
and $a_{2}=1$ (with $a_{min}<1$).
According to the restriction (\ref{restricao}), the potential $V(a)$ gives us an oscillatory solution for the scale factor (between $a_{min}\leq a \leq 1$) avoiding the singularity formation at the center of the matter distribution at $a=0$.
\par
Let us now consider the following coordinates transformation
\begin{eqnarray}
\label{eqIII.25}
R=ar~,~~\bar{\theta}=\theta~,~~\bar{\phi}=\phi .
\end{eqnarray}
In this sense, the line element (\ref{eq1}) can be written as
\begin{eqnarray}
\label{eqIII.26}
ds^2=-\Big[1-\frac{\dot{a}^2R^2}{a^2-kR^2} \Big]dt^2-\frac{2a\dot{a}R}{a^2-kR^2}dtdR+\frac{a^2}{a^2-kR^2}dR^2+R^2d\bar{\Omega}^2,
\end{eqnarray}
where $d\bar{\Omega}^2\equiv d{\Omega}^2$.
Defining
\begin{eqnarray}
\label{eqIII.27}
\nonumber
T:=F[S(R,t)],~~C(R,t):=\Big[1-\frac{\dot{a}^2R^2}{a^2-kR^2} \Big],\\
\nonumber
E(R,t):=-\frac{a\dot{a}R}{a^2-kR^2}~,~D(R,t):=\frac{a^2}{a^2-kR^2},
\end{eqnarray}
we get:
\begin{eqnarray}
\label{eqIII.28}
\nonumber
ds^2=-C \Big(\frac {\partial S}{\partial t}\Big)^{-2} \Big(\frac {d F}{d S}\Big)^{-2} dT^2+2\Big(\frac {\partial S}{\partial t}\Big)^{-1}\Big(\frac {d
F}{d S}\Big)^{-1}
%\times
\Big[ C \Big(\frac {\partial S}{\partial t}\Big)^{-1} \Big(\frac {\partial S}{\partial R}\Big)+E \Big]dTdR\\
-\Big[ C \Big(\frac {\partial S}{\partial t}\Big)^{-2} \Big(\frac {\partial S}{\partial R}\Big)^{2} -D + 2E\Big(\frac {\partial S}{\partial t}\Big)^{-1} \Big(\frac
{\partial S}{\partial R}\Big)\Big]dR^2+R^2d{\Omega}^2.~~~~~
\end{eqnarray}
\par
According to the Birkhoff theorem\cite{weinberg} in General Relativity, we know that the exterior
solution of a spherically symmetric collapse of dust is given by the Schwarzschild geometry where the metric is diagonal. As we are motivated to find a correction of the Schwarzschild geometry, let us consider the following condition
\begin{eqnarray}
\label{eqIII.29}
C \Big(\frac {\partial S}{\partial t}\Big)^{-1} \Big(\frac {\partial S}{\partial R}\Big)+E=0.
\end{eqnarray}
Therefore the metric (\ref{eqIII.28}) is diagonal and we automatically guarantee that
\begin{eqnarray}
\label{eqIII.30}
g_{RR}=\Big[1-\frac{R^2}{a^2}(k+\dot{a}^2)\Big]^{-1}.
\end{eqnarray}
It is easy to verify
that the solution for (\ref{eqIII.29}) is given by
\begin{eqnarray}
\label{eq7}
S(R,t)=\delta+\frac{\mu}{a}\sqrt{a^2-kR^2}\exp{\Big(-k\int \frac{1}{a\dot{a}^2}da\Big)}
\end{eqnarray}
where $\delta$ and $\mu$ are arbitrary constants,
\begin{eqnarray}
\label{eqIII.37}
\nonumber
{-k\int{\frac{1}{a\dot{a}^2}}da}=D\ln|a-1|-I\ln|a-a_{min}|~~~~~~~~~~~~~~\\
+\frac{2J}{\sqrt{4B-A^2}}\arctan{\Big(\frac{2a+A}{\sqrt{4B-A^2}}\Big)}
\end{eqnarray}
with
\begin{eqnarray}
\label{coeficientes}
\nonumber
A&:=&1+a_{min}-\frac{8\pi G_{N}E_{0}}{3k}~~,~~B:=\frac{4\pi G_{N}E^2_{0}}{3k|\sigma|a_{min}}~~,\\
\nonumber
D&:=&\frac{1}{(1-a_{min})(A+B+1)}~~,~~~~~~~~~~\\
\nonumber
I&:=&\frac{a^3_{min}}{(1-a_{min})(a^2_{min}+Aa_{min}+B)}~,\\
\nonumber
J&:=&\frac{B(Da_{min}+E)}{a_{min}}.~~~~~~~~~~~~~~~~~~~~~~
\end{eqnarray}
\par
Let us now assume that the junction of the interior solution with the exterior geometry is given at the surface
defined by $r=\gamma$, where $\gamma$ determines the boundary of the matter distribution.
By defining the constant $M:=\frac{4\pi \gamma^3 E_{0}}{3}$, we obtain
\begin{eqnarray}
\label{eqIII.30a}
g_{RR}=\Big[1-\frac{2G_{N}M}{R}+\frac{3G_{N}M^2}{4\pi|\sigma|R^4}\Big]^{-1}\Big|_{r=\gamma}.
\end{eqnarray}
\par
Employing the integrating factor technique,\cite{weinberg} we define the function $F[S(R,t)]$ by the following differential equation
\begin{eqnarray}
\label{FuncaoF}
\frac{d F} {d S}= \frac{a^2 {\dot{a}}}{a^2-R^2(k+{\dot{a}}^2)}\Big[ \frac{\sqrt{a^2-kR^2}}
{k(S-\delta)} \Big],
\end{eqnarray}
where $t$ and $R$ are implicit functions of $S$ through equation (\ref{eq7}).
Adopting this choice we verify from equation (\ref{eqIII.28}) that
\begin{eqnarray}
\label{eqIII.30new}
g_{TT}=-\Big[1-\frac{R^2}{a^2}(k+\dot{a}^2)\Big]=-\frac{1}{g_{RR}}.
\end{eqnarray}
\par
It is important to remark that when $r=\gamma$, $S(a)$ is a monotonous function in the physical domain of $a$,\footnote{In order to illustrate the behavior of the integral (\ref{eqIII.37}) in the physical
domain of our parameters, let us assume $|\sigma|=0.05m^{-2}$, $G_{N}=1$, $E_{0}=1.04\times10^{-23}{m^{-2}}$,
 $\gamma=6.96\times10^8m$, $\delta=1$, $\mu=0.01$. We are assuming a reasonable physical value for the brane tension
while considering a density one hundred times greater than the density of the sun.
By taking these parameters we see that $S(a)$ is a monotonous function of $a$ in the physical domain $a_{min}=4.702669449\times10^{-8}\leq a\leq a_{max}=1$.} which may be properly inverted in such a way that we can express $a$ in terms of $S$.
\par
Therefore, from equations (\ref{eqIII.30a}) and (\ref{eqIII.30new}) we see that the exterior geometry reads
\begin{eqnarray}
\label{eq1se}
\nonumber
ds^2=-\Big[1-\frac{2G_{N} M}{R}+\frac{3G_{N} M^2}{4\pi|\sigma|R^4}\Big]dT^2
+\Big[1-\frac{2G_{N} M}{R}+\frac{3G_{N} M^2}{4\pi|\sigma|R^4}\Big]^{-1}dR^2~~~~~~~~~~\\
+R^2(d{\theta}^2+\sin^2{{\theta}}d{\phi}^2).~~~~~~
\end{eqnarray}
\par
From (\ref{eq2}) and (\ref{eq1se}) we can see that General Relativiy is recovered, with
the corresponding standard Oppenheimer-Snyder models, when $|\sigma|\rightarrow\infty$\cite{oppen,weinberg} .
\par
In the following Sections we discuss some properties of the exterior geometry (\ref{eq1se})
and construct its maximal analytical extension in order to better understand the avoidance of
the Schwarzschild singularity. We also discuss some physical issues in this geometry as the
Hawking temperature of the black hole and an extension of  black hole thermodynamics in this spacetime.

\section{Analytic Completion of the Manifold}

${}$
In order to examine the analytic completion of the exterior geometry (\ref{eq1se}), we need to
know whether, and under what circumstances, the configuration forms event horizons.
By defining the polynomial $P(R):=R^4 g_{TT}$, we see that
a necessary condition for horizon formation is that the mass $M$, of the collapsing star, equals or exceeds
a critical limit $M_{\ast}$~,
\begin{eqnarray}
\label{eq11}
M \geq M_{\ast} \equiv \Big( \frac{4}{9 \pi G_N^3~|\sigma|}\Big)^{1/2}~,
\end{eqnarray}
($|\sigma|$ fixed), namely, that the polynomial $P(R)$ has
one ${R}_{\ast}$, or two (${R}_{-} < {R}_{+}$) roots respectively (cf. Fig. 1).
Otherwise we cannot have formation of event horizons.
For illustration let us consider again the parameters given in (a) but now with
a large value of the brane tension ($|\sigma|=10^8$) to tentatively approach general relativity.
It is easy to check that we obtain again a strictly
monotonous function S(a) in the physical range of a, and $R_{-}>\gamma a_{min}$ -- specifically
$\gamma a_{min} = 0.02597827765$ and $R_{-} = 0.02597828531$. As $|\sigma|$ decreases, the difference
($R_{-}-\gamma a_{min}$) increases by the same order of magnitude.
\par
On the other hand, we note that the critical mass depends solely on the parameter $|\sigma|$. To have
an idea of the order of magnitude of $|\sigma|$ for event horizon formation, let us take
$M_{\ast}\sim 1.4 M_{\odot}$, the Chandrasekhar limit. This yields $|\sigma| \sim 10^{-2} km^{-2}$. A star with
the Chandrasekhar mass will not form an event horizon if the brane tension is
smaller than $10^{-2} km^{-2}$, so that this value establishes a lower bound for $|\sigma|$.
\par
Considering then the case of two positive
real roots, the polynomial $P(R)$ may be rewritten as
\begin{eqnarray}
\label{eqP}
P(R) = (R-R_{+})(R-R_{-})(R^2+\alpha R+\beta),
\end{eqnarray}
where
\begin{eqnarray}
\label{eqP1}
\alpha=R_{+}+R_{-}-2G_{N}M ~,~~ \beta=\frac{3G_{N}M^2}{4\pi|\sigma|R_{+}R_{-}}.
\end{eqnarray}
\begin{figure}
\begin{center}
{\includegraphics*[height=6.5cm,width=7.5cm]{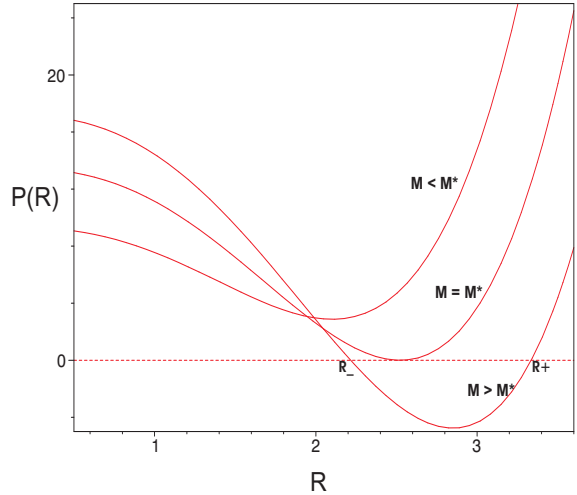}}
\caption{The polynomial
$P(R) \equiv R^4g_{TT}$ for dust masses $M<M_{\ast}$
(no event horizon), $M=M_{\ast}$ (extremal black hole) and $M>M_{\ast}$ (black hole formation with a event horizon ${R}_{+}$ and a Cauchy horizon ${R}_{-}$). We consider here $|\sigma|=0.05$, in units $G_N =c=1$.}
\label{Fig. 1}
\end{center}
\end{figure}
In this case the collapse of the surface of
dust must cross ${R}_{-}$~, that is, $\gamma a_{m}< {R}_{-}$~ so that a {\it stable} black hole
forms with trapped perpetually bouncing matter.
\par
Let us then consider the following coordinates transformation
\begin{eqnarray}
\label{eqi1}
\frac{2\kappa}{u}du:=\frac{R^4}{P(R)}dR-dT,
\end{eqnarray}
\begin{eqnarray}
\label{eqi2}
\frac{2\kappa}{v}dv:=\frac{R^4}{P(R)}dR+dT.
\end{eqnarray}
Therefore, from (\ref{eq1se}) we get
\begin{eqnarray}
\label{eqi3}
\nonumber
ds^2=\frac{4\kappa^2}{uv}\frac{P(R)}{R^4}dudv
+R^2(d{\theta}^2+\sin^2{{\theta}}d{\phi}^2).
\end{eqnarray}
By defining
\begin{eqnarray}
\label{eqi4}
R^{\ast}:=\int g_{RR} dR,
\end{eqnarray}
\begin{figure}
\begin{center}
{\includegraphics*[height=6cm,width=12cm]{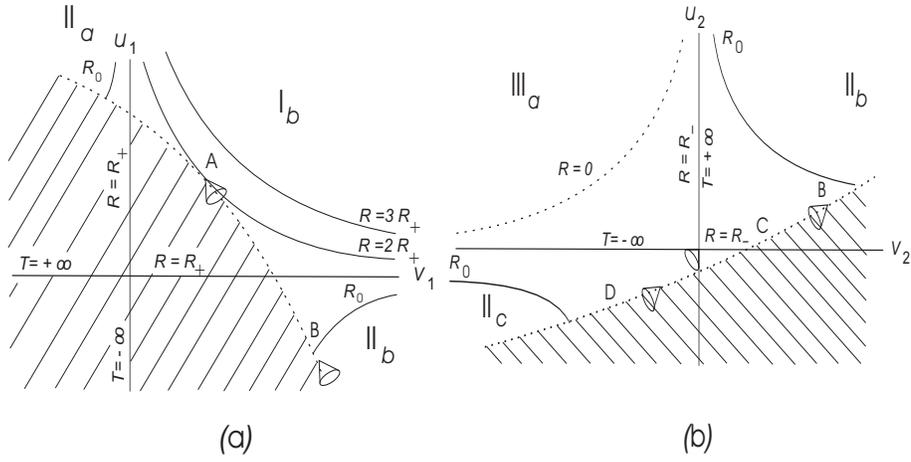}}
\caption{Kruskal diagrams showing the curve ABCD which illustrates how the matter surface evolves in time. The shaded portion corresponds to the interior of the matter distribution. Figures (a) and (b) overlap in the region $II_{b}$, and may be regarded as linked together along the curve $R=R_{0}$, where $R_{0}$ is any convenient value between $R_{+}$ and $R_{-}$.
}
\label{Fig. 2}
\end{center}
\end{figure}
we get that
\begin{eqnarray}
\label{eqi5}
\nonumber
R^{\ast}=R+G\ln{\Big|\frac{R}{R_{+}}-1\Big|}+W\ln{\Big|\frac{R}{R_{-}}-1\Big|}~~~~~~~~~~~~\\
+\frac{2J}{\sqrt{4\beta-\alpha^2}}
\arctan{\Big(\frac{2R+\alpha}{\sqrt{4\beta-\alpha^2}}\Big)},
\end{eqnarray}
where
\begin{eqnarray}
\label{eqi6}
G&:=&\frac{R^4_{+}}{(R_{+}-R_{-})(R^2_{+}+\alpha R_{+}+\beta)}~~,~~\\
W&:=&-\frac{R^4_{-}}{(R_{+}-R_{-})(R^2_{-}+\alpha R_{-}+\beta)}~~,\\
J&:=&\beta\Big(\frac{A}{R_{+}}+\frac{B}{R_{-}}-1\Big).~~
\end{eqnarray}
Therefore, integration of (\ref{eqi1}) and (\ref{eqi2}) yields
\begin{eqnarray}
\label{eqi7}
R^{\ast}=\kappa\ln|uv|~~,~~T=\kappa\ln|v/u|.
\end{eqnarray}
\par
Consider now the chart $(u_1,v_1)$ (defined for $R>R_{-}$) obtained by setting $\kappa=G$. From (\ref{eqi5}) and (\ref{eqi6}) we find that
\begin{eqnarray}
\label{eqi8}
\nonumber
u_{1}v_{1}=\Big(\frac{R}{R_{+}}-1\Big)\Big(\frac{R}{R_{-}}-1\Big)^{-\frac{|W|}{G}}~~~~~~~~~~~~~~~~~~~~~~~~~~~~\\
\times\exp{\Big[\frac{R}{G}+
\frac{2J}{G\sqrt{4\beta-\alpha^2}}
\arctan{\Big(\frac{2R+\alpha}{\sqrt{4\beta-\alpha^2}}\Big)}\Big]}.
\end{eqnarray}
and (\ref{eqi3}) exhibits no singularity at $R=R_{+}$. The chart $(u_{1},v_{1})$ in fact gives a regular mapping
of any given subregion of the manifold which has $R>R_{-}$. However, a coordinate singularity does develop at $R=R_{-}$ and it is necessary to go over another chart before that happens.
\par
Define the chart $(u_2,v_2)$ (defined for $R<R_{+}$) by setting $\kappa=B$. From (\ref{eqi5}) and (\ref{eqi6}) we find that
\begin{eqnarray}
\label{eqi9}
\nonumber
u_{2}v_{2}=\Big(1-\frac{R}{R_{+}}\Big)^{-\frac{G}{|W|}}\Big(\frac{R}{R_{-}}-1\Big)~~~~~~~~~~~~~~~~~~~~~~~~~~~\\
\times\exp{\Big[\frac{R}{W}+
\frac{2J}{W\sqrt{4\beta-\alpha^2}}
\arctan{\Big(\frac{2R+\alpha}{\sqrt{4\beta-\alpha^2}}\Big)}\Big]},
\end{eqnarray}
and this provides a regular covering for any subregion with $R<R_{+}$.
\par
In the domain of overlap $R_{-}<R<R_{+}$ the two charts are related by
\begin{eqnarray}
\label{eqi10}
|u|^G_{1}=|u|^W_{2}~~,~~|v|^G_{1}=|v|^W_{2}.
\end{eqnarray}
\par
Figures 2(a), 2(b) are Kruskal-type diagrams which together give a faithful map of any subregion covered by a pair of overlapping charts $(u_1,v_1)$ and $(u_2,v_2)$.
The maximal analytical extension of spacetime (\ref{eq1se}) is analogous
to that of a Reissner-Nordstr\"om black hole\cite{chandra,israel2} with an exterior event horizon $R_{+}$
and a Cauchy horizon $R_{-}$ (cf. Fig. 3).
\begin{figure}
\begin{center}
{\includegraphics*[height=10cm,width=7cm]{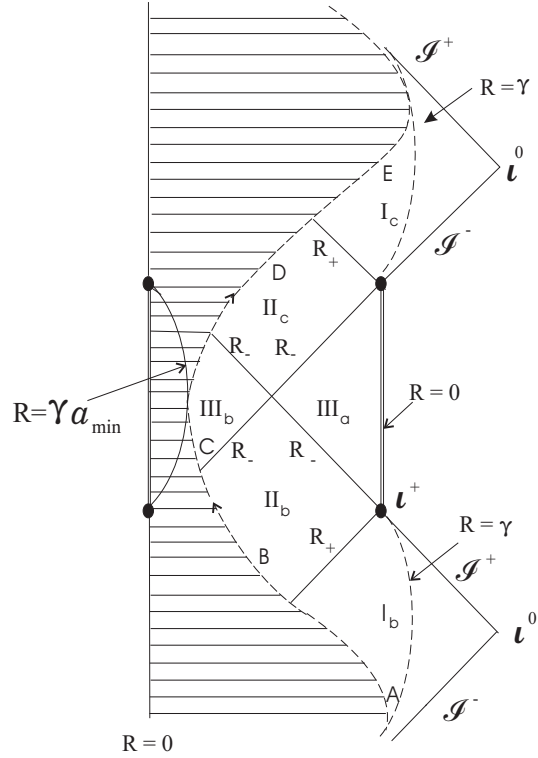}}
\caption{Penrose diagram for the spacetime assuming $M > M_{\ast}$. The infinite chain of asymptotically flat regions I $(\infty>R>R_{+})$ are connected to regions III $(R_{-}>R>\gamma a_{min})$ by regions II $(R_{+}>R>R_{-})$. The shaded portion, limited by $R=\gamma a_{min}$ and
$R=\gamma$ corresponds to the interior of the matter distribution.
The same dotted line ABCDE illustrates how the matter surface evolves in time. Once crossed $R_{-}$ such a surface keeps bouncing between $R=\gamma$ and $R=\gamma a_{min}$.}
\label{Fig. 3}
\end{center}
\end{figure}
Here the singularity in the interior of the matter distribution is barred by the timelike surface $R=\gamma a_{m}$
where the spacetime extension ends.
\par
By taking $M_{\ast}$ as the maximum value for a white dwarf to be in equilibrium
($1.4M_{\bigodot}$), we have that the domain for the brane tension is given by the lower limit
\begin{eqnarray}
\label{eq34r2}
\nonumber
|\sigma| \gtrsim 10^{-8}  \frac{1}{m^2}.
\end{eqnarray}
In fact, the behavior of the $S$ function does not change when
$|\sigma|>10^{-7}\frac{1}{m^2}$. Therefore we justify the typical value $|\sigma|=0.05$ adopted in our numerical illustration in Sec. 2.
\section{The Hawking Temperature}
In 1975, S. W. Hawking derived -- through a semi-classical approach -- the thermal spectrum of emitted particles by a black hole.\cite{hawking} In this Section we are going to follow this same original procedure in order to derive the corrections in the Hawking radiation.
\par
Let us consider a massless Klein-Gordon field $\varphi$ in the background defined by the exterior spacetime (\ref{eq1se}).
The propagation of such scalar test field is taken to be governed by the scalar wave equation
\begin{eqnarray}
\label{eqKGN}
g^{ab}\nabla_a\nabla_b ~\varphi=0.
\end{eqnarray}
Exploiting the symmetries of the background
we seek a solution as
\begin{eqnarray}
\label{eq38}
\varphi_{\omega ml}=\frac{1}{R} R_{\omega l}(R^{\ast}) Y_{ml}(\theta, \phi) \exp (-i\omega t).
\end{eqnarray}
Substituting this expression for $\varphi$, the wave equation is reduced to an ordinary differential equation
in $R^{\ast}$ for the modes $R_{\omega l}$ and given by
\begin{eqnarray}
\label{eq41}
\nonumber
\frac{d^2R_{\omega l}}{d{R^{\ast}}^2}+\Big\{ \omega^2 -\frac{1}{R^2} \Big[ l(l+1) + \frac{2G_N M}{R}-\frac{3G_N M^2}{\pi |\sigma|R^4}\Big]~~~~~~~~~~~~~~~~\\
\Big(1-\frac{2G_N M}{R}+\frac{3G_N M^2}{4\pi |\sigma|R^4}\Big)\Big\}R_{\omega l}=0.
\end{eqnarray}
As $r\rightarrow\pm\infty$ we have that
\begin{eqnarray}
\label{eq42}
\nonumber
\frac{d^2R_{\omega l}}{d{R^{\ast}}^2}+ \omega^2R_{\omega l}=0 \Rightarrow R_{\omega l}(R^{\ast})=\exp(\pm i\omega{R^{\ast}})
\end{eqnarray}
and, asymptotically, one can express the Klein Gordon field as
\begin{eqnarray}
\label{eq43}
\varphi_{1}=\frac{1}{R} \exp[-i\omega(t-{R^{\ast}})] Y_{ml}(\theta, \phi)
\end{eqnarray}
and
\begin{eqnarray}
\label{eq44}
\varphi_{2}=\frac{1}{R} \exp[-i\omega(t+{R^{\ast}})] Y_{ml}(\theta, \phi).
\end{eqnarray}
\par
Let us now assume that the source that generates the exterior solution (\ref{eq1se}) is given by a thin shell of a spherically symmetric matter distribution, and the flat spacetime inside such distribution is given by
\begin{eqnarray}
\label{eq46}
ds^2=-dt^2+dr^2+r^2d\Omega^2.
\end{eqnarray}
Defining $a(t)$ as the scale factor that the describes the evolution of the matter distribution, we impose that the interior metric match the exterior geometry by the following equation
\begin{eqnarray}
\label{eq50}
\nonumber
1-\Big(\frac{da}{dt}\Big)^2=\Big(1-\frac{2G_N M}{a}+\frac{3G_N M^2}{4\pi |\sigma|a^4}\Big)\Big(\frac{dT}{dt}\Big)^2~~~~~~~~~~~~~~~~~~\\
-\Big(1-\frac{2G_N M}{a}+\frac{3G_N M^2}{4\pi |\sigma|a^4}\Big)^{-1}\Big(\frac{da}{dt}\Big)^2.
\end{eqnarray}
We also define the respective null interior and exterior coordinates by
\begin{eqnarray}
\label{eq47}
V:=t+r~,~~U:=t-r~,
\end{eqnarray}
and
\begin{eqnarray}
\label{eq48}
v:=T+R^{\ast}~,~~u:=T-R^{\ast}.
\end{eqnarray}
Let us now assume that the null incident rays get into the matter distribution when $a=a_I\gg R_{+} \sim GM$. Therefore we have that
\begin{eqnarray}
\label{eq51}
\Big(1-\frac{2G_N M}{a_I}+\frac{3G_N M^2}{4\pi |\sigma|{a_I}^4}\Big)\rightarrow1.
\end{eqnarray}
and
\begin{eqnarray}
\label{eq52}
\Big(\frac{dT}{dt}\Big)^2\simeq 1\rightarrow t\simeq T.
\end{eqnarray}
On the other hand,
\begin{eqnarray}
\label{eq53}
v_I\simeq t+R^{\ast}\Rightarrow V_I=v_I+\kappa
\end{eqnarray}
where
\begin{eqnarray}
\label{eq54}
\kappa=a_I-R^{\ast}(a_I).
\end{eqnarray}
\par
When $r=0$, we derive the trivial relation between $V$ and $U$ at the center of the matter distribution:
\begin{eqnarray}
\label{eq55}
V_0=t=U_0.
\end{eqnarray}
\par
Let us now consider that the outgoing waves emerge from the matter distribution when $a=a_{II}\sim R_{+}$. If
$t_{0}$ is taken to be the instant in which $a=R_{+}$, one may expand the scale factor $a_{II}(t)$ in Taylor series as
\begin{eqnarray}
\label{eq56}
a_{II}(t)\simeq R_{+} + F(t_{0}-t),
\end{eqnarray}
where F is a constant.
\par
Therefore, from equation (\ref{eq50}) we have
\begin{eqnarray}
\label{eq57}
T\simeq - \lambda\ln(t_{0}-t)
\end{eqnarray}
up to first order in $(t_{0}-t)$, where
\begin{eqnarray}
\label{eq571}
\lambda\equiv \frac{\sqrt{2G_N MR^3_{+}-3G_N M^2/4\pi|\sigma|}}{4R_{+}-6G_N M}.
\end{eqnarray}
However, from (\ref{eqi5}) we have
\begin{eqnarray}
\label{eq58}
R^{\ast}\simeq G\ln{\Big(\frac{a}{R_{+}}-1\Big)}.
\end{eqnarray}
Then we get
\begin{eqnarray}
\label{eq59}
\nonumber
u_{II}\simeq-\delta\ln{(t_{0}-t)},
\end{eqnarray}
where $\delta\equiv\lambda+G$. Therefore,
\begin{eqnarray}
\label{eq60}
U_{II}\simeq \chi \exp{\Big(-\frac{u_{II}}{\delta}\Big)}+\psi
\end{eqnarray}
where
\begin{eqnarray}
\label{eq61}
\chi=-(1+F)~,~~\psi=t_{0}-R_{+}.
\end{eqnarray}
However, at the origin of the coordinate system we have $U_0=V_0$. Therefore the relation between the exterior null coordinates is given by
\begin{eqnarray}
\label{eq63}
v=v_{0}+\chi\exp{\Big(-\frac{u}{\delta}\Big)}~,~~u=-\delta\ln{\Big(\frac{v-v_{0}}{\chi}\Big)}
\end{eqnarray}
where
\begin{eqnarray}
\label{eq64}
v_{0}:=\psi-\kappa.
\end{eqnarray}
\par
Using (\ref{eq44}) we expand $\varphi_{1\omega lm}$ in terms of $\varphi_{2\omega lm}$ as
\begin{eqnarray}
\label{eq641}
\nonumber
\varphi_{1\omega lm}= \int^\infty_0 [\alpha^{\ast}_{\omega^{\prime} \omega l m} \exp{(-i\omega^{\prime} v)}
-\beta_{\omega^{\prime} \omega l m} \exp{(i\omega^{\prime} v)}]d\omega^{\prime},
\end{eqnarray}
where $\alpha^{\ast}_{\omega^{\prime} \omega l m}$ and $\beta_{\omega^{\prime} \omega l m}$ are the Bogolubov coefficients.\cite{birrel} Therefore,
it is straightforward to show\cite{hawking} that
\begin{eqnarray}
\label{eq644}
|\alpha_{\omega^{\prime} \omega l m}|=\pi \omega \delta |\beta_{\omega^{\prime} \omega l m}|.
\end{eqnarray}
However, it follows from the orthogonality propriety of $\varphi_{1\omega lm}$ and $\varphi_{2\omega lm}$ that
\begin{eqnarray}
\label{eq645}
\sum_{\omega^{\prime}} [~|\alpha_{\omega^{\prime} \omega l m}|^2 - |\beta_{\omega^{\prime} \omega l m}|^2~]=1.
\end{eqnarray}
Therefore we obtain that the spectrum of the average number of created particles on the $\omega lm$ mode is given by
\begin{eqnarray}
\label{eq41rr}
N_{\omega lm}=\sum_{\omega^{\prime}} |\beta_{\omega^{\prime} \omega l m}|^2 =\frac{1}{\exp{(2\pi\delta\omega)}-1}.
\end{eqnarray}
The above result corresponds to a Planckian spectrum with associated temperature
\begin{eqnarray}
\label{eq42rr}
T_H=\frac{\Big(R_{+}-R_{-}\Big)}{2 \pi R_{+}~\zeta} ,
\end{eqnarray}
where
\begin{eqnarray}
\zeta=\frac{R_{+}^{3}}{(R_{+}^{2}+ \alpha R_{+}+\beta)}+\Big[\frac{\sqrt{3G_{N} M ~(2R_{+}^{3}- M/4\pi|\sigma|)}(R_{+}-R_{-})}{{ 2R_{+}(2R_{+}-3G_{N} M)}}
\Big].
\end{eqnarray}
\par
The
Hawking temperature depends on the parameters $M$ and $|\sigma|$.
We note that in the extremal case we have $R_{+} = R_{-} =
3G_{N}M/2$ implying that $T_{H}\rightarrow 0$ continuously as $R_{+}\rightarrow
R_{-}$.
The observation of Hawking radiation could, in principle,
allows us to test our results for finite $|\sigma|$.
Another feature, which demands a carefully analysis, is related to the entropy.
We dedicate Sec. 5 to this subject.

\section{Black Hole Thermodynamics and the Quest for a Statistical Mechanics Model for the Entropy of Quasi-extremal Black Holes}
${}$
\par Motivated by the analysis of energy processes involving black holes Bekenstein\cite{bekenstein}
made the remarkable assumption that the entropy of a black hole should be proportional to the area of its event horizon
and formulated a First Law of Black Hole Thermodynamics where the surface gravity of the black hole appeared
as proportional (via dimensional fundamental constants) to a temperature.
Bekenstein's results however did not involve any fundamental principle of statistical mechanics.
Two years later S. W. Hawking,\cite{hawking} by examining the quantum creation of particles
near a Schwarzschild black hole, showed that the black hole emits particles with a Planckian thermal spectrum of temperature
$T_{BH}=\kappa /2 \pi$ (in units $\hbar=c=G_N =K_{B}=1$) where $\kappa=1/4 M$ is the surface gravity of the black hole.
This striking result fits exactly in the Bekenstein formula for the First Law of black hole thermodynamics,
thus validating Bekenstein's proposals and fixing the proportionality factor connecting the entropy and the area of the black hole.
\par Nevertheless by considering the classical theory of General Relativity, the singularity issue
still posed an insurmountable barrier to the task of constructing a model for the interior of the black hole
and of counting its degrees of freedom, what would eventually lead to a definition of entropy.
\par
The results of the geometrical Black Hole Thermodynamics of Bekenstein are recovered in our model
for the case of quasi-extremal black holes with an additional term connected to the work done by the variation of the brane tension. Let us consider a small deviation from the extremal case,
with $R_{\pm}=\frac{3G_{N} M}{2}\pm \epsilon$ where $\epsilon$ is infinitesimal. Neglecting higher order terms
in $\epsilon$ we have from $P(R_{\pm})=0$ that
\begin{eqnarray}
\label{eq0eerb}
\epsilon=\sqrt{\frac{3G^2_{N}M^2}{8}-\frac{1}{6 \pi G_{N}|\sigma|}},
\end{eqnarray}
from which the useful relation is derived,
\begin{eqnarray}
\label{eq0eerb-1}
d \epsilon \simeq  \frac{1}{2 \epsilon} \Big(\frac{3 G_{N}^2 M_{\ast}}{4}~ dM + \frac{1}{6 \pi G_N |\sigma_{\ast}|^2} ~d|\sigma|  \Big),
\end{eqnarray}
and which is valid about the extremal configuration. The condition for horizon formation is given by $\epsilon\geq0$.
\begin{figure}
\begin{center}
{\includegraphics*[height=5cm,width=6cm]{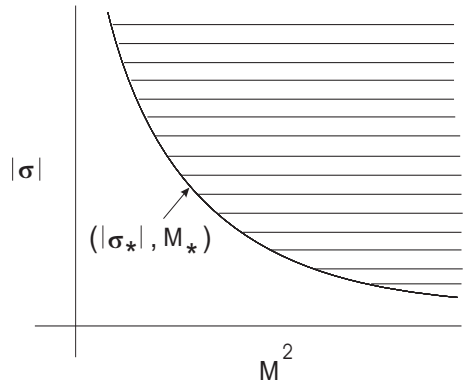}}
\caption{The parameter space ($|\sigma|, M^2$): the shaded area
corresponds to black hole configurations with two horizons; the
white area corresponds to configurations with no black hole formation.
The condition $\epsilon=0$ defines the limiting curve ($|\sigma_{\ast}|, M_{\ast}$)
that corresponds to extremal black hole configurations. The region that we are considering here is given by
sufficiently large values of $|\sigma|$, and very near the extremal curve on the shaded side.}
\label{Fig4}
\end{center}
\end{figure}
The equality in this latter relation defines a curve (cf. Fig. 4) in the parameter space ($|\sigma|, M^2$)
that corresponds to extremal black holes. The region above the curve is the region of black holes (with two horizons)
while the region below defines configurations with no horizon formation and therefore no black holes.
In this sense, $\epsilon$ corresponds to a small deviation from the curve ($|\sigma_{\ast}|, M_{\ast}$) towards the
black hole area of the parameter space. In this approximation the Hawking temperature is small and given by
\begin{eqnarray}
\label{eqth2}
T_{H}\simeq \Big(\frac{8\hbar}{9\pi K_{B}G^2_{N}M^2_{\ast}}\Big)~\epsilon,
\end{eqnarray}
where we now restore the constants $\hbar$ and $K_{B}$.
By defining the the outer horizon area as $A_{\rm outer}:=4\pi R^2_{+} \simeq \Big( A_{\rm extr}+12 \pi G_N M_{\ast} \epsilon\Big)$, we obtain for the quasi-extremal case that
\begin{eqnarray}
\label{eqth3}
\frac{K_{B}}{4G_{N}\hbar}dA_{\rm outer}\simeq \frac{1}{T_{H}} \Big(dM+\frac{M_{\ast}}{2|\sigma_{\ast}|}d|\sigma|\Big).
\end{eqnarray}
where Eq. (\ref{eq0eerb-1}) has been used. We can therefore associate the
horizon area of the quasi-extremal black hole with the geometrical entropy
\begin{eqnarray}
\label{eqGEO}
S_{geom}=\frac{K_{B}}{4G_{N}\hbar}A_{\rm outer},
\end{eqnarray}
a result which is in accordance to Bekenstein's definition.\cite{bekenstein} Equation (\ref{eqth3}) is an extended First Law with
an extra work term connected to the variation of the brane tension. For deviations with $|\sigma|={\rm const.}$
we recover the form of the First Law for the Schwarzschild black hole in the brane scenario.
\par
We are now led to tentatively construct a statistical mechanics model for the bouncing collapsed
matter (in the quasi-extremal case) with an associated partition function engendering a thermodynamics which, under certain assumptions,
may be connected to the geometrical thermodynamics discussed above. The following facts about
the quasi-extremal configurations are fundamental to our approach. The quasi-extremal configurations are basically characterized
by the parameter $\epsilon$ (cf. Eq. (\ref{eq0eerb})) that fixes not only the oscillatory motion of the collapsed matter but also
all the properties of the extended spacetime, in particular determining the geometrical thermodynamic variables of the exterior spacetime.
Actually $\epsilon$ is a measure of all the fluctuations about the extremal case, either occurring in the dynamics of
the oscillating collapsed matter or determining the horizon fluctuations of the exterior geometry.
\par To proceed let us consider the dynamical equation for the scale factor expressed by the constraint
(\ref{eq2}), with the initial conditions $\dot{a}(0)=0$ and $a(0)=1$.
Once the surface of matter distribution is defined by $R\equiv\gamma a$, we can define the momentum (per unity of mass) at the surface distribution as
$p_{R}:=\gamma~ \dot{a}$ so that the dynamical equation for the scale factor is given by the Hamiltonian constraint
\begin{eqnarray}
\label{eq4eerb}
H=\Big[\frac{p^2_{R}}{2}+V(R)\Big]=0,
\end{eqnarray}
where
\begin{eqnarray}
\nonumber
\label{eq5eerb}
V(R):=\frac{3G_{N}M^2}{8 \pi |\sigma|R^4}-\frac{G_{N}M}{R}+\frac{k \gamma^2}{2}.
\end{eqnarray}
Expanding (\ref{eq4eerb}) in a neighborhood of $R=R_{\ast}=3 G_N M_{\ast}/2$ (the extremal configuration) we obtain
\begin{eqnarray}
\label{eq6.neerb}
\mathcal{H}:=\frac{p^2_{R}}{2}+\frac{4}{9G^2_{N}M_{\ast}^2}\Big[\frac{8}{3G_{N}M_{\ast}}\epsilon^2R
+\Big(1-\frac{40}{9G^2_{N}M^2_{\ast}}\epsilon^2\Big)R^2\Big]\simeq - V(R_{\ast}).~~~~
\end{eqnarray}
As the brane formulation must approach General Relativity in the low energy limit, it is natural to expect the value
of $|\sigma_{\ast}|$ to be sufficiently large. In this instance the brane tension satisfies the inequality
$\sqrt{1/(3\pi G_{N}|\sigma_{\ast}|)}\ll 1$, which implies that the term proportional to $R$ in (\ref{eq6.neerb}) can be neglected.
The constraint (\ref{eq6.neerb}) is then approximately given by
\begin{eqnarray}
\label{eq7eerb}
H=\frac{p^2_{R}}{2}+\frac{1}{2}\omega^2R^2 \simeq - V(R_{\ast}),
\end{eqnarray}
where
\begin{eqnarray}
\label{eq8eerb}
\omega\equiv\frac{2\sqrt{2}}{3G_{N}M_{\ast}}\sqrt{\Big(1-\frac{40}{9G^2_{N}M_{\ast}^2}\epsilon^2\Big)}\simeq \frac{2\sqrt{2}}{3G_{N}M_{\ast}}-
\frac{40\sqrt{2}}{27G^3_{N}M_{\ast}^3}\epsilon^2.
\end{eqnarray}
\par
As we are assuming a non-interacting fluid, the interior particles of the matter distribution must also oscillate with a frequency $\omega$.
We should note that the first term in the second equality of (\ref{eq8eerb}) corresponds to the oscillation
of the matter distribution in the extremal case, which is thermodynamically a configuration of zero temperature, while the second term corresponds
to oscillations generated by small deviations from the extremal configuration which depend on the parameter $\epsilon$. This same parameter
is responsible for horizon fluctuations about the extremal case which give rise to the geometrical Hawking temperature
associated with the exterior spacetime, as seen from (\ref{eqth2}). In analogy with the result (\ref{eqth2}) of the Hawking temperature,
we make the provisional assumption that the second term in (\ref{eq8eerb}) arises from fluctuations which we denote {\it thermal}, connected to a temperature $T_{S}$ given by
\begin{eqnarray}
\label{eqts}
T_{S}:=a ~\Big(\frac{8\hbar}{9\pi K_{B} G^2_{N}M^2_{\ast}}\Big)~ \epsilon,
\end{eqnarray}
where $a$ is an adimensional constant. In this sense, from the point of view of statistical mechanics,
the extremal configuration would obviously have a zero partition function, being a zero temperature configuration.
However this is not the case in the quasi-extremal cases for which a partition function can be constructed by
quantizing the {\it thermal} fluctuations appearing (\ref{eq8eerb}) as we proceed to show.
\par Let us define $N_{\ast}$ as the number of Planck masses contained in the matter of the extremal case, $N_{\ast}=M_{\ast}/m_{\rm Pl}$.
The approximated motion of our system can then be interpreted as the 1-dim motion of $N_{\ast}$ noninteracting oscillators with frequency $\omega$,
the energy levels of which -- under a quantization procedure -- will be given by $E_n= (n+1/2) \hbar \omega$. The fluctuations about
the extremal configuration present in $\omega$ and now parametrized with the temperature $T_S$ will engender quantum {\it thermal}
fluctuations that will have a fundamental contribution in the partition function.
\par
The canonical partition function of the system may then be expressed as
\begin{eqnarray}
\label{eq9eerb}
\nonumber
Z=\Big\{\sum^{\infty}_{n=0}\exp\Big[-\Big(n+\frac{1}{2}\Big)\beta\hbar\omega\Big]\Big\}^{N_{\ast}}
=\Big[\frac{\exp(\beta\omega\hbar/2)}{\exp(\beta\omega\hbar)-1}\Big]^{N_{\ast}} \\
\simeq {\rm exp}(-N_{\ast}\beta\omega\hbar/2),~
\end{eqnarray}
where $\beta\equiv 1/K_{B}T_{S}$ and the third equality results from $T_S$ being small for the quasi-extremal case.
\par The free energy is given by
\begin{eqnarray}
\label{eq11eerb}
F=-{\cal R}~T_{S}\ln{Z}\simeq\frac{N_{\ast}{\cal R}\omega\hbar}{2K_{B}},
\end{eqnarray}
where ${\cal R}$ is an appropriate constant. By definition the entropy of the system can be calculated as a function of the free energy through the relation
$S=-{\partial F}/{\partial T_{S}}$, resulting in
\begin{eqnarray}
\label{eq12eerb}
S_{stat} \simeq \frac{40\sqrt{2}N_{\ast}{\cal R}\hbar}{27K_{B}G^3_{N}M^3_{\ast} a^2} T_{S}
 = \frac{5 \sqrt{2}\pi N_{\ast} {\cal R}}{3 G_N M_{\ast} a}~\epsilon,
\end{eqnarray}
where use was made of Eq. (\ref{eqts}).
Therefore, using the relations (\ref{eq0eerb})-(\ref{eq0eerb-1}), we obtain
\begin{eqnarray}
\label{eq14eerb-11}
dS_{stat}\simeq \frac{1}{T_{S}}\Big(dM+\frac{M_{\ast}}{2|\sigma_{\ast}|}d|\sigma|\Big),
\end{eqnarray}
where we have fixed  ${\cal R}:=K_{B}\tilde{N}$, with $\tilde{N}=\frac{9 \sqrt{2}}{10}N_{\ast} \simeq 1.272 N_{\ast}$. We see that $\tilde{N}$
arises naturally as the analog of an Avogadro number for the internal matter distribution of the extremal case.
Also Eq. (\ref{eq14eerb-11}), which is a first law of thermodynamics for the bouncing collapsed matter, validates our definition
(\ref{eqts}) of $T_S$ as a temperature.
\par By comparing Eqs. (\ref{eqth3}) and (\ref{eq14eerb-11}) we are led to identify $T_S$ with the Hawking temperature $T_H$ which was derived in Section 4
for the exterior spacetime, therefore fixing $a=1$ in (\ref{eqts}). We should remark that the statistical entropy derived in (\ref{eq12eerb})
differs from Bekenstein's entropy $S_{geom}$ (\ref{eqGEO}) by a zero temperature additive constant which corresponds to the
area of the extremal black hole about which our treatment is made, namely
\begin{eqnarray}
\label{eqw}
S_{geom}=\frac{K_{B}}{4\hbar G_{N}}A_{extr}+S_{stat}~.
\end{eqnarray}
In fact we can see that
the statistical entropy (\ref{eq12eerb}) arises just from the quantum {\it thermal} fluctuations with which the partition function is built,
in distinction to the geometrical entropy of Bekenstein.
\par Notwithstanding these striking similarities, a question that can now be posed is how the concepts of Hawking temperature and Bekenstein entropy
can be connected with the above defined entropy and temperature of the interior bouncing matter. In fact,
in accordance with the geometrical black hole thermodynamics, one might argue that the entropy of the black hole is an
external variable connected to the event horizon boundary of the exterior gravitational field. Therefore the entropy of the collapsed matter would thus be
irrelevant for physical processes outside the black hole. However, as far as quasi-extremal configurations are considered, we can
suggest a mechanism of how the quantum {\it thermal} fluctuations may connect the entropy of the collapsed matter to fluctuations of
the outer event horizon. To see this, let us consider the fundamental frequency $\Delta \omega$, of a particle
of the oscillating matter distribution with mass $m_{pl}$, given by (cf. (\ref{eq8eerb}))
\begin{eqnarray}
\label{eqnew1}
\Delta \omega=\frac{40\sqrt{2}}{27G^3_{N}M_{\ast}^3}\epsilon^2.
\end{eqnarray}
Its associated momentum is given by $\Delta p= \sqrt{2 \hbar m_{pl} \Delta \omega}$. According to Heisenberg uncertainty principle
the uncertainty in its localization $\Delta R \geq \hbar/2$ and, using (\ref{eqnew1}), we obtain
\begin{eqnarray}
\label{eqnew2}
\Delta R \geq \frac{1}{2\epsilon} \sqrt{\frac{27G^3_{N}M_{\ast}^3\hbar}{40\sqrt{2}~m_{Pl}}}.
\end{eqnarray}
For illustration, by taking $M_{\ast}$ as the Chandrasekhar mass ($M_{\ast}=1.4 M_{\odot}$) we find $\Delta R\geq 0.4127626961\times 10^{-14}m^2/\epsilon$.
On the other hand $R_{+}\sim 3G_{N} M_{\ast}/2=3\times 10^3m$. Therefore, if $\epsilon \lesssim 10^{-18} m$, namely
smaller than $10^{17}l_{P}$ (where $l_{P}$ is the Planck length), we can assure that the scale of the fluctuations $\Delta R > R_{+}$
so that the quantum {\it thermal} fluctuations that give rise to the entropy (\ref{eq12eerb}) of the collapsed matter might be connected
to the fluctuations of the event horizon in the quasi-extremal case.
In this sense, Eqs. (\ref{eqth3}) and (\ref{eq14eerb-11}) validate the identification of both entropies (up to an additive constant corresponding to the area of the extremal black hole) with the proviso that the temperature responsible for the {\it thermal} fluctuations is identified with the Hawking temperature.
\par
An extension of our analysis beyond the quasi-extremal approximation will be the subject of a future work. In this instance
the collapsed matter distribution should be strongly correlated and its description as noninteracting oscillating particles
should be drastically modified.
\section{Conclusions and Final Comments}
${}$
\par
In this paper we have examined the gravitational collapse of pure dust in a braneworld scenario. By considering a timelike extra dimension
of the bulk we show that the classical singularity produced by the collapse can be avoided, resulting instead in a bouncing nonsingular solution.
For the case of the total mass $M$ of the collapsing dust greater than a critical value $M_{\ast}$ (which depends on the brane tension)
the smooth junction of this solution with the exterior geometry results in a spacetime analogous to the Reissner-Nordstr\"om geometry, with
an exterior event horizon and a Cauchy horizon which enclose the collapsed bouncing matter. We construct the analytical extension of the spacetime
up to the surface of the bouncing collapsed matter where the two geometries are smoothly connected. For $M=M_{\ast}$  we have the extremal black hole
configuration with one horizon only. This case plays a fundamental role in our discussion of possible entropy definitions for the system.
\par
In the case of $M > M_{\ast}$, the effects of the bulk-brane corrections on the classical observational tests of General Relativity were calculated but not included
in the paper. However for completeness some of these results are briefly commented in the following. To start let us note that
the dynamics of General Relativity is recovered in the limit $|\sigma| \rightarrow \infty$.
Concerning the advance of the planetary perihelion, due to the brane-bulk corrections in the geometry,
the perihelion precession per revolution decreases when compared to the predicition of General Relativity,
for the present case of a timelike extra dimension. On the contrary, for a spacelike extra dimension, the perihelion
precession increases. In both cases the observational data\cite{perihelio} restrict the brane tension $|\sigma|$
to be greater than $10^{-8} {m^{-2}}$.
\par\
Analogously for the bending of light rays passing in the neighborhood of a spherical massive body, the
effects predicted by General Relativity are attenuated due to the corrections in the case of a timelike
extra dimension. In the case of a spacelike extra dimension the deflection angle of the asymptotes increases.
It is worth remarking that the observational data\cite{sobral} do not impose a new limit for the brane tension.
In fact, in this case we only need that $|\sigma| \geq 10^{-12}m^{-2}$ for both cases of a timelike or spacelike extra dimension.
Summarizing the effects of the bulk-brane corrections on the classical tests of General Relativity are attenuated (increased) when one considers a
timelike (spacelike) extra dimension. Although the predictions of General Relativity are corroborated by the observational tests in
the solar system, the corrections of our model may turn out to be significative when we consider higher energy and/or curvature scales.
\par The central result of our work is contained in Secs. $4$ and $5$. In Sec. 4 we calculated (through a semiclassical approach) of the corrected Hawking temperature
in our model. We also obtain that, analogous to the case of General Relativity, we have the prediction of a zero temperature for the case of extremal black holes.
The calculation of the modified Hawking temperature allowed us to derive, analogously to Bekenstein, a geometric entropy
that confirms (for quasi-extremal black holes) the classical prediction that the entropy is proportional to the area of the event horizon, as shown
in the first part of Sec. $5$.
Although the classical black hole thermodynamics introduced by Bekenstein was validated by Hawking's semiclassical derivation of
the black body thermal emission of a black hole, with a temperature proportional the the gravity surface of the event horizon of the
black hole, black hole thermodynamics always seemed to possess a heuristic character since no basic principle of statistical mechanics was used
in its derivation. Indeed, the definition of the entropy of black holes is still an open problem and we actually refer to it as a geometrical entropy.
\par
In this context, in the remaining of Sec. $5$, we have tentatively constructed a simple statistical mechanics model for the entropy of the oscillatory collapsed matter
in the case of quasi-extremal black holes. In fact, by quantizing the internal degrees of freedom of the quasi-extremal configuration, we
constructed a statistical partition function and a free energy -- and assuming that the temperature responsible for the {\it thermal}
fluctuations obeys the same form of the Hawking temperature -- we derive an entropy that satisfies a first law of thermodynamics
strikingly similar to the geometrical first law for the quasi-extremal black hole. This statistical entropy differs from Bekenstein's entropy
by a zero temperature additive constant which corresponds to the area of the extremal black hole about which our treatment is made.
In fact we can see that the statistical entropy arises from the quantum {\it thermal} fluctuations with which the partition function is built,
in distinction to the geometrical entropy of Bekenstein.
\par
We should finally mention that the presence of the Cauchy horizon $R_{-}$ in the maximal analytical extension
of the geometry (\ref{eq1se}) may pose the question of a possible instability of the spacetime as measured by geodetic
observers.\cite{chandra} We are aware that this may constitute a problem to our statistical treatment of the
bouncing collapsed matter, but not to the geometrical entropy formation of Sec. 6 for the exterior spacetime.
We intend to approach this issue in a future publication.

\section{Acknowledgements}

The authors acknowledge the partial financial support of CNPq/MCTI-Brazil.

\end{document}